# Evolution of Spatial and Multicarrier Scheduling: Towards Multi-cell Scenario

Pol Henarejos[1]*, Ana Perez-Neira[1,2], Velio Tralli[3], Marco Moretti[4], Nikos Dimitriou[5], and Giulio Dainelli[5]

[1]Centre Tecnòlogic de Telecomunicacions de Catalunya (CTTC), Barcelona, Spain
`{pol.henarejos,ana.perez}@cttc.es`
[2]Universitat Politècnica de Catalunya (UPC), Barcelona, Spain,
[3]CNIT, University of Ferrara, Ferrara, Italy,
`velio.tralli@unife.it`
[4]Dipartimento di Ingegneria dell'Informazione, Universita di Pisa, Italy,
`marco.moretti@iet.unipi.it`
[5]Institute of Acelerating Systems & Applications, National Kapodistrian University of Athens, Athens, Greece,
`nikodim@phys.uoa.gr`

**Abstract.** OFDMA systems are considered as the promising multiple access scheme of next generation multi-cellular wireless systems. In order to ensure the optimum usage of radio resources, OFDMA radio resource management algorithms have to maximize the allocated power and rate of the different subchannels to the users taking also into account the generated co-channel interference between neighboring cells, which affects the received Quality of Service. This paper discusses various schemes for power distribution schemes in multiple co-channel cells. These schemes include centralized and distributed solutions, which may involve various degrees of complexity and related overhead and may employ procedures such as linear programming. Finally, the paper introduces a new solution that uses a network flow model to solve the maximization of the multi-cell system sum rate. The application of spatial beamforming at each cell is suggested in order to better cope with interference.

## 1 Introduction

Most of the existing literature on resource allocation focuses on the single cell scenario, where all users are assigned to a different portion of the available spectrum. However, mobile communication systems are better described as multi-cellular systems where either coordinated or uncoordinated cells transmit on the same bandwidth and are therefore the capacity is increased, but not unaffected by Multiple Access Interference (MAI). MAI in particular deteriorates the performance of users near cell boundaries. Thus, any resource allocation problem in a multi-cell environment has to take into account the impact of the MAI on the system. A frequency reuse factor larger than one guarantees a large reduction of the interference at the cost of a reduction of the efficiency in the usage of spectral resources. For this reason in recent literature several works have focused on Orthogonal Frequency Division Multiple Access (OFDMA) allocation in multi-carrier cellular systems with a frequency reuse factor equal to one. Due to the strong impact of MAI in this scenario, it is important to take full advantage of frequency and multi-user diversity of the system. The authors from [1] set the initial point to start the current research, in which is addressed this manuscript.

---

* This work was supported by the European Commission under the Network of Excellence NEWCOM++ (216715). The authors thank the partners of WPR8 in Newcom++ NoE for their valuable technical help.

In the past years, several approaches relying on the concept of inter-cell coordination have emerged, which can be distinguished in two categories: packet-based coordination and resource-allocation based coordination. In the first one, data packets destined to the users are replicated at several base stations, before jointly precoding/beamforming, and transmitting from all the Base Station antennas (BS) [2], [3], [4]. The drawback of this approach is a large overhead in inter-cell signaling, packet routing, and feedback for exchanging the channel state information required to compute the precoders. In the second approach, the interference is tackled by means of coordinated resource control (power, scheduling, etc.) between the cells [5], [6] which make lower complexity and distributed coordination techniques are possible. Power control and smart soft reuse partitioning are possible strategies that can be applied [7], [8], [9].

Dynamic multi-cell power control targeted at maximizing the sum of user rates in the network is a very difficult task and does not lend itself easily to a distributed (across the cells) implementation, except for some particular cases with a large number of users [10]. The reason is as follows: dynamic power control affects the Signal to Interference plus Noise Ratio (SINR) of all users in all cells in a fully coupled manner making interference unpredictable.

OFDMA resource allocation is a viable solution to exploit channel and multi-user diversity in wireless communication systems. In a multi-user scenario, with an OFDMA multiple access scheme, each user is assigned a subset of orthogonal subcarriers. If the transmitter has full knowledge of the Channel State Information (CSI), subcarriers can be assigned with the goal of maximizing some optimality criterion. Since the radio propagation channels are statistically independent among the users, what is a bad channel for one user may be a good one for another and thus, thanks to the effect of multi-user diversity, dynamic resource allocation largely increases the system spectral efficiency.

Resource Allocation schemes in a multi-cell scenario can be divided into centralized and distributed algorithms. Centralized schemes perform allocation through a central unit like the Radio Network Controller (RNC) that collects CSI and interference level for each user in the system. Ideally, the RNC decides which subchannels (or subcarriers) to assign to each single user with the suitable format and power level. On the other hand, in a distributed algorithm resource allocation is performed autonomously by each single BS in its cell. The main problem for centralized schemes is the large amount of signaling needed for exchanging CSI and allocation feedback. Moreover, the allocation complexity grows exponentially with the number of users in the network, since resource assignment is realized by a single unit, which has to process large amounts of data. Thus, centralized algorithms are often studied to provide an ideal bound for the performance of others schemes. Distributed algorithms require lower complexity and signaling, since they perform allocation locally at each BS and therefore require only the information about the users in the cell. However, such solutions very often lead to iterative algorithms, which may have convergence problems. Sometimes the differences between distributed and centralized schemes blur away since distributed schemes may require a limited amount of centralized information to improve their performance.

Recently, the spatial diversity has been introduced to reach an acceptable performance. Many solutions of the current State of Art (SoA) propose Multiple Input Multiple Output (MIMO) techniques, widely extended for the single cell scenarios. Nevertheless, these schemes treat inter-cell interference as noise, where the performance is limited, specially for edge-cell users. The authors in [11], [12] and [13] deal with this kind of problem. Unfortunately, computational power and complexity raise up with the number of cells. Thus, distribution forms of cooperation among the user terminals and BS appear with great interest. One alternative is cooperative MIMO to minimize total power with QoS constraints [14]. This scheme implies that BS are going to change their peak power and may be not suitable in several scenarios. This manuscript present a second alternative of cooperative MIMO to maximize a cost function of rate. If the cost function is the sum rate function, the problem becomes NP-hard [15]. However, other cost functions are possible to decrease its complexity [16].

This document aims to organize the SoA according to the requirements of the system, if the interference management is relevant or not. Moreover, two algorithms to distribute users in the several cells and power allocation are also presented. Thereby, section 2 introduces the different techniques used in the OFDMA power distribution in multi-cell scenarios. Section 3 presents a scheme to distribute users in the several cells, with centralized or distributed complexity, based on the Linear Programming (LP). Furthermore, section 4 describes a framework to perform the power allocation and user selection from the approach of a single-cell multi-antenna scenario. Finally, this document is ended by the conclusions.

## 2 Power distribution for OFDMA multi-cell systems

Power has an important role in multi-cell systems not only for the rate optimization, but for the interference management. Networks where interference is not a big problem (such as those where there is frequency planning and adjacent cells use different OFDMA subcarriers) may strive to optimize the throughput. On the contrary, in networks where interference plays an important aspect, power strategy may be oriented to limit this interference. The following algorithms cover both categories.

**2.1 Layered and distributed dynamic resource allocation algorithm**

A downlink communication system in a cellular network, where all cells adopt a frequency reuse factor equal to one, is considered. Each user CSI and the level of interference on each subcarrier are assumed to be perfectly known by each BS. Since in each cell a subchannel is allocated to at most one user, the effects of MAI depend on the users (and specifically on their location and power/rate allocation) that are allocated the same channel in adjacent cells. The MAI on channel $m$ affecting user $k$ in cell $q$ is:

$$I_{k,m,q} = \sum_{j=1, j \neq q}^{Q} p_{m,q} G_{k,m,j} \tag{1}$$

where $p_{m,q}$ indicates the power transmitted by cell $q$ on subcarrier $m$ and $G_{k,m,j} = |h_{k,m,q}|^2$ is the channel gain between user $k$ and cell $q$ on subchannel $m$. Thus, the power required for achieving a certain target SINR is:

$$p_{k,m,q} = SINR \frac{BN_0 + I_{k,m,q}}{G_{k,m,q}} \tag{2}$$

where $B$ stands for the bandwidth of the signal and $N_0$ is the noise power in $W/Hz$.

A layered architecture that integrates in each cell a Packet Scheduler (PS) with an adaptive resource allocator (RA) is considered. First, the RA allocates the resources with the goal of minimizing the transmitted power in each cell subject to user's rate constraints to keep low the MAI. To exploit multi-user diversity, the RA tends to assign most of the resources to the users that have good channel condition.

Second, the PS enforces long-term fairness in order to compensate the short term displacement of resources due to the RA. Moreover, a load control mechanism is introduced to force the convergence of the allocation. To reduce the complexity of the allocation phase, all users adopt only one transmission format with spectral efficiency $\eta_0$ for all the subcarriers so that rate constraints are translated into a number of resources, i.e., $R_{k,q} = B\eta_0 m_{k,q}$, where $m_{k,q}$ stands for the number of channels allocated to user $k$ at cell $q$. The cost of a resource for a given user is the power required for achieving spectral efficiency $\eta_0$.

Whenever a cell modifies its allocation, it changes also the interference experienced by users in neighboring cells, which in turn change their allocation. Thus, the allocation phase is iterated until a stable allocation is reached in all cells. To help convergence, if the system is not able to reach a stable allocation, the load is progressively reduced in all cells.

After a stable allocation is reached, the PS updates the maximum rate requirements $m_{k,q}$ for each user in each cell with the goal of achieving long-term fairness. The convergence is not always guaranteed but the combined actions of load control and packet scheduling push towards convergence and fairness at the same time. An additional action to ensure convergence is provided by a mechanism where the most power consuming users can be progressively switched off.

This approach tends to be unfair because users near the cell boundary risk to be too penalized in terms of the reduction of the available bandwidth. If not carefully designed, the main drawback of this scheme is the number of iterations required for achieving a stable allocation.

**Minimum feedback layered and distributed dynamic resource allocation algorithm** A minimum feedback scheduling technique extends the concepts of distributed allocation that was outlined in the previous Section. It is assumed that each user measures the interference of each subcarrier and sends to the BS only the interference values that correspond to the "best" subcarriers. The number of those "best" subcarriers, i.e. those which experience the lowest MAI, is a parameter to be determined by the system operator. For the other subcarriers, the RA algorithm assumes the worst case scenario and assigns to them a fixed high value of interference. With this approach the required feedback is reduced while the RA algorithm has still the necessary inputs in order to be able to provide results. Of course, in this case the allocations will not be the optimal since the algorithm is forced to work without the actual interference values for all the subcarriers. However, it can be argued that it may provide similar throughput results as the previous algorithm with much less overhead and complexity [17].

**Random subcarrier allocation algorithm** Another possible way to allocate resources to the users consists of a random resource allocation to the users in each cell, without any kind of optimization criteria. In this case the PS sets again the maximum number of subcarriers which can be assigned to the users, and then the allocator assigns randomly the subcarriers with respect to the constraints set by the PS. In this case, after random allocation, only power control takes place and after that the subcarriers which have not achieved their SINR target are switched off.

Additionally, users in the outer region of a cell will use a portion of the bandwidth which is different from that one utilized by the users in the outer region of the adjacent cells.

**2.2 Power planning**

The objective is to achieve a fully distributed implementation of resource allocation over a multi-cell OFDMA network, whose aim is minimizing the network outage capacity. To reach this goal, the selection of the user to be scheduled and of the resources (here defined as the couple subcarrier/transmit power level) to be assigned to him, should be performed taking into account the channel gain and the received interference power. If a fully distributed approach is pursued, each BS can only rely on local information provided via a feedback channel by its own set of users. So, in this work some structuring inside the system is introduced, in order to make interference level inside the network predictable.

Though in principle power levels can continuously vary inside a predefined range, only a certain set of possible power levels are assumed, and these are distributed among cells and subcarriers according to a predefined pattern. This concept will be denoted from now on as "power planning".

In particular, the network is organized in groups of $Q$ adjacent cells according to a regular pattern as done for frequency planning in 2G systems and, for analogy, this group of cells is denoted as "cluster". Then, the $M$ equally spaced OFDMA subcarriers assigned to each cell are arranged in $Q$ groups of $M/Q$ adjacent subcarriers, from now on denoted also as "sub-bands". It is clear that the larger the value of $Q$, the smaller the frequency diversity if correlation between subcarriers is taken into account.

A power vector $\mathbf{P} = [p_1 \ldots p_Q]$ is introduced, which is composed of the $Q$ power levels, also denoted as elements of the "power profile". Hence, in the allocation process only these $Q$ power values are usable. From now on, this vector will be denoted as "multi-cell transmit power vector". Thus, the terms "power profile" and "multi-cell transmit power vector" are considered to represent identical things.

In each cell, every sub-band is assigned with one of the values belonging to power vector $\mathbf{P}$, and over all sub-bands inside a cell all values of $\mathbf{P}$ are exploited. Nevertheless, looking at a specific sub-band, the set of cells belonging to the same cluster use all power levels available in $\mathbf{P}$.

So, each cell in the network is assigned with a tag $j$ ranging from 1 to $Q$ denoting the cell type. Then, since each tag is assigned with a specific power vector (i.e., with a specific order of the possible $Q$ power levels in vector $\mathbf{P}$), cells with the same tag will be assigned with the same power vector, whereas cells belonging to the same cluster are assigned with permutations of the original power vector.

## 3 Multi-cell user assignment

Previous section was addressed to manage the power budget and choose which strategy can result more effective. However, this aspect also separates the power variable from the others. One of the advantages of decoupling the power variable depending on scenario requirements is the fact that users can be scheduled a posteriori following the same requirements. Even though the power can be used to schedule users, i.e. power equal to zero implies no user is scheduled, LP methods has been considered to be an efficient tool to solve this kind of problems. On the contrary, since power and user scheduling is performed in separated steps, solution becomes suboptimal.

Authors in [18] present a resource allocator for the uplink of multi-cell OFDMA systems. That concept is also applicable for the downlink channel. Previous section has defined the power strategy and power is solved in this point. Hence, user scheduling can be performed through LP. Even though authors in [18] minimize the power, the maximization of sum rate can also be pursued. Moreover, LP offers the chance to introduce more variables to the problem to make it as so general as it is desired

Consider a downlink OFDMA system with $Q$ cells with one BS each, $K$ users distributed over all cells and $M$ carriers available in each BS. Since power is defined in the previous sections, all rates of users are pre-defined in the following manner. The rate of $k$th user at $m$th carrier and $q$th cell is:

$$r_{k,m,q} = \log\left(1 + \frac{G_{k,m,q}p_{m,q}}{\sum_{q'\neq q}^{Q} G_{k,m,q'}p_{m,q'} + BN_0}\right) \qquad (3)$$

where $p_{m,q}$ is the power served by $q$th cell at carrier $m$, defined previously.

This scheme can be easily combined with beamforming to decrease the effect of the interference. Thus, the BS may use a beam to increase the SINR and the overall sum rate.

### 3.1 Centralized algorithm

A centralized approach could be done by modeling the multi-cell system as a single network with one central control unit which computes the access parameters for all users in all cells and the way with which users are scheduled over the network. In order to assign users to cells and carriers, LP algorithms are used and they prove to be a good way to exploit this kind of scenarios.

The problem can be stated as:

$$\begin{aligned}\mathbf{b} &= \arg\max_{\mathbf{b}} \sum_{q=1}^{Q}\sum_{k=1}^{K}\sum_{m=1}^{M} b_{k,m,q} r_{k,m,q} \\ s.t. &\sum_{k=1}^{K} b_{k,m,q} \leq 1, \ q=1,\ldots,Q, \ m=1,\ldots,M \\ &\sum_{q=1}^{Q} b_{k,m,q} \leq 1, \ k=1,\ldots,K, \ m=1,\ldots,M\end{aligned} \qquad (4)$$

where $b_{k,m,q} = 1$ if user $k$ is scheduled at $m$th carrier and in the $q$th cell and 0 otherwise. The first constraint implies that only one user can be scheduled in each carrier and cell. The second constraint, implies that only one cell can be assigned to one user at same carrier. This problem can be solved by LP easily. However, it requires centralized schemes and feedback corresponding to all $h_{k,m,q}$ that must be known at the transmitters. Although complexity is proportional to each variable that is introduced, results are near optimal.

### 3.2 Distributed algorithm

The centralized algorithm is optimal compared to the distributed algorithm since it has more information about the channels of all users in all cells. On the other hand, having a centralized algorithm requires a huge amount of feedback information processing complexity and introduces large amounts of overhead in the calculations. The idea is to distribute the complexity in each cell, i.e. removing $q$ index. Hence, each BS should execute the following algorithm separately.

A distributed algorithm could be derived from the above as:

$$\mathbf{b} = \arg\max_{\mathbf{b}} \sum_{k=1}^{K} \sum_{m=1}^{M} b_{k,m,q} r_{k,m,q}$$
$$s.t. \sum_{k=1}^{K} b_{k,m,q} \leq 1, \ m = 1, \ldots, M. \tag{5}$$

Note that the second constraint is removed since it requires a centralized way of controlling all users scheduled in all cells. Thereby, one user can be scheduled in different cells at the same carrier.

This simplification distributes complexity over the network and does not require any centralized processing. Additionally, the amount of feedback can be reduced if interference is assumed to be equal to all users in all cells. That is equivalent to approximate the rate of $k$th user at $m$th carrier and $q$th cell as:

$$r_{k,m,q} = \log\left(1 + \frac{G_{k,m,q} p_{m,q}}{I_{m,q} + BN_0}\right). \tag{6}$$

It is easy to show that the $q$th cell only requires channel gains $G_{k,m,q}$ of its $K$ users at each carrier.

## 4 From spatial to multi-cell scheduling

In [19] the authors present a spatial scheduler for multicarrier systems in a single-cell scenario. The basis is a network flow formulation for maximization of the system sum rate. To sumarize it, the spatial diversity is solved using Multiuser Opportunistic Beamforming and choosing the user permutation, and its corresponding beam set, that achieve the best sum rate; then, the power allocation is performed from this spatial allocation. This section proposes a modification of the algorithm in [19] and distributes the spatial dimension, separating the antennas one from each others and distributing one per BS. For this reason, instead of beam-user selection, cell-user selection has to be carried out. The work in [19] considers ergodic sum rate maximization for continuous rates. The ergodic framework also allows the optimization in the time domain. In fact, if $[1, \ldots, N]$ is the time interval of the optimization, for any generic system or user metric, $R[n]$, under ergodic assumption for random processes in the system, the approximation $(1/N)\sum_n R[n] \approx \mathbb{E}\{R[n]\} = \mathbb{E}\{R\} = \mathcal{R}$ holds, where $\mathcal{R}$ does not depend on time $n$. Hence, optimizing $\mathcal{R}$ means optimizing $R[n]$ over time interval $[1, \ldots, N]$.

The discrete variable or index $u_{m,q} \in \mathbb{K}_0 = \{0, 1, \ldots, K\}$ indicates the user (i.e. 0 means no user) that is scheduled to use cell $q$ on subcarrier $m$. Note that only one user or none can be scheduled for each carrier and each cell. The whole set of these variables is the matrix $\mathbf{U} \in \mathbb{K}_0^{M \times Q}$, whereas the whole set of powers is the matrix $\mathbf{P} \in \mathbb{R}^{+,M \times Q} \cup \{0\}$. It is implicitly assumed that if $u_{m,q} = 0$ then $p_{m,q} = 0$[2].

The aim of resource allocation is to dynamically assign radio interface resources to the different users, i.e. to determine optimal values of $\mathbf{U}$ and $\mathbf{P}$. The problem can be formulated as

$$\max_{\mathbf{U},\mathbf{P}} \sum_{k=1}^{K} \mathcal{R}_k(\mathbf{U},\mathbf{P})$$
$$s.t.\ \mathcal{P}_q(\mathbf{U},\mathbf{P}) \leq \bar{\mathcal{P}},\ \forall q \quad (7)$$
$$\mathcal{R}_k(\mathbf{U},\mathbf{P}) \geq \phi_k \sum_{s=1}^{K} \mathcal{R}_s(\mathbf{U},\mathbf{P}),\ \forall k$$

where $\mathcal{R}_k(\mathbf{U},\mathbf{P}) = \mathbb{E}\{R_k(\mathbf{U},\mathbf{P})\} = \sum_{m=1}^{M} \sum_{q=1}^{Q} \mathbb{E}\{\delta_k^{u_{m,q}} r_{k,m,q}\}$ is the rate provided to user $k$ from (3), $\mathcal{P}_q(\mathbf{U},\mathbf{P}) = \sum_{m=1}^{M} \mathbb{E}\{p_{m,q}\}$ is the total average power spent by cell $q$ to serve the allocated users and $\delta_k^u$ is the Kronecker's delta[3]. The first constraint refers to the total power used which must be less than a maximum amount $\bar{\mathcal{P}}$. The second constraint implies that users ought to obtain the proportional $\phi_k$ part of the sum rate, which determines the share of throughput finally achieved by each user. Therefore, $\boldsymbol{\phi}$ must satisfy the condition $\sum_{k=1}^{K} \phi_k = 1$.

It is important to underline that in this problem rate and power constraints are referred to as average values. In this way, the instantaneous constraints are relaxed leading to a reduction in the complexity of the resulting optimization algorithm.

### 4.1 Dual optimization framework and adaptive algorithms

The optimization problem is non convex and Lagrangian duality [20] is used to solve the problem. It enables each user to adapt their resources locally with the aid of limited information exchange. An interesting point of the Lagrangian is the dual decomposition into individual user and cell terms. This fact motivates decentralized algorithms as in [20] and allows to distribute the complexity over the network. However, to obtain a distributed algorithm as seen later, it is necessary to decouple user assignment from power assignment. The dual objective of problem (7) is defined as

$$\min_{\boldsymbol{\lambda} > \mathbf{0}, \boldsymbol{\mu} \geq \mathbf{0}} g(\boldsymbol{\lambda}, \boldsymbol{\mu}) = \min_{\boldsymbol{\lambda} > \mathbf{0}, \boldsymbol{\mu} \geq \mathbf{0}} \left\{ \max_{\mathbf{U},\mathbf{P}} \mathcal{L}(\mathbf{U},\mathbf{P},\boldsymbol{\lambda},\boldsymbol{\mu}) \right\} = \min_{\boldsymbol{\lambda} > \mathbf{0}, \boldsymbol{\mu} \geq \mathbf{0}} \mathcal{L}(\mathbf{U}^*,\mathbf{P}^*,\boldsymbol{\lambda},\boldsymbol{\mu}) \quad (8)$$

---

[2] This also means that $\mathbf{P}$ has an implicit dependence on $\mathbf{U}$ and vice versa as shown afterwards.

[3] $\delta_k^u = 1$ if $u = k$ and 0 otherwise.

where $\mathcal{L}(\mathbf{U}, \mathbf{P}, \boldsymbol{\lambda}, \boldsymbol{\mu})$ is the Lagrangian function of the problem (7) and $\boldsymbol{\lambda}, \boldsymbol{\mu}$ are the Lagrangian multipliers.

It is important to remark that while the primal problem is a non-concave maximization, the dual problem becomes a convex optimization. However, the dual problem is not differentiable and an iterative subgradient method is used to update the $K + Q$ solutions of the dual problem at each discrete time instant. Starting from initial solutions $\boldsymbol{\lambda}^0$ and $\boldsymbol{\mu}^0$, the update equations at the $i$th iteration derive from subgradient expressions and are:

$$\begin{aligned} \boldsymbol{\lambda}^{i+1} &= \left[\boldsymbol{\lambda}^i - \delta_{\boldsymbol{\lambda}} \left(\bar{\mathcal{P}}_q - \mathcal{P}_q(\mathbf{U}^{*i}, \mathbf{P}^{*i})\right)\right]_{\epsilon}^{+} \\ \boldsymbol{\mu}^{i+1} &= \left[\boldsymbol{\mu}^i - \delta_{\boldsymbol{\mu}} \left(\mathcal{R}_k(\mathbf{U}^{*i}, \mathbf{P}^{*i}) - \phi_k \sum_{s=1}^{K} \mathcal{R}_s(\mathbf{U}^{*i}, \mathbf{P}^{*i})\right)\right]^{+} \end{aligned} \qquad (9)$$

where $[x]_{\epsilon}^{+} = \max(\epsilon, x)$, $0 < \epsilon \ll 1$ and $\delta_{\boldsymbol{\lambda}}, \delta_{\boldsymbol{\mu}}$ are positive step-size parameters. $\mathbf{U}^{*i}, \mathbf{P}^{*i}$ indicate the optimal solutions of the Lagrangian at the $i$th iteration, i.e. those which maximize $\mathcal{L}(\mathbf{U}, \mathbf{P}, \boldsymbol{\lambda}^i, \boldsymbol{\mu}^i)$.

### 4.2 Solutions for the allocation problem

The optimal power and user solutions are difficult in that case due to the cross dependence of user and power allocation. Therefore, the dual objective can be rewritten as follows:

$$g(\boldsymbol{\lambda}, \boldsymbol{\mu}) = \max_{\mathbf{U}, \mathbf{P}} \mathcal{L}\left(\mathbf{U}, \mathbf{P}, \boldsymbol{\lambda}, \boldsymbol{\mu}\right) = \sum_{q=1}^{Q} \lambda_q \bar{\mathcal{P}} + M \mathbb{E}\left\{\max_{\mathbf{u}_m} \left[\max_{\mathbf{p}_m \geq \mathbf{0}} \mathcal{M}(\mathbf{u}_m, \mathbf{p}_m)\right]\right\} \qquad (10)$$

with

$$\mathcal{M}(\mathbf{u}_m, \mathbf{p}_m) = \sum_{q=1, u_{m,q} \neq 0}^{Q} \left[(\mu_{u_{m,q}} - \boldsymbol{\mu}^T \boldsymbol{\phi}) \log_2(1 + r_{u_{m,q}, m, q}(\mathbf{p}_m)) - \lambda_q p_{m,q}\right]. \qquad (11)$$

The optimal solution, given $\boldsymbol{\lambda}, \boldsymbol{\mu}$, becomes, for each frequency $m$,

$$\mathbf{u}_m^* = \arg\max_{\mathbf{u}_m} \mathcal{M}^*(\mathbf{u}_m) \qquad (12)$$

with

$$\mathcal{M}^*(\mathbf{u}_m) = \max_{\mathbf{p}_m \geq \mathbf{0}} \mathcal{M}(\mathbf{u}_m, \mathbf{p}_m). \qquad (13)$$

This shows that user selection (12) and power allocation (13) are decoupled from the dual optimization and both them can be computed separately. In fact, user selection is computed before the power solution is found.

Concerning spatial allocation, the main issue is to reduce the search space. This issue can be faced by using suboptimal greedy selection procedures. The simplest among them is the opportunistic selection. Thereby, each user selects the best cell by assuming that all base stations are transmitting with a preassigned power and feeds back the selected cell with its SINR, while each cell allocates its resources to the best user selected among those competing for that cell. This can be done helped by spatial beamforming at each base station. Next sub-section comments further on that.

### 4.3 Discussion on Centralized and Distributed solutions

The optimal solution requires a centralized controller that runs all or parts of the algorithms. Even though the power allocation algorithm based on the update of $\boldsymbol{\lambda}^i$ can be distributed on each base station, user allocation algorithm requires a centralized solution, i.e. a controller which knows all channel gains determines, for all subcarriers, the vector $\mathbf{u}_m$ and send it to base stations through signaling.

A decentralized implementation can be set up by using the opportunistic suboptimal solution of power allocation. In this case user allocation algorithm has two steps (for each subcarrier):

– Each user selects the best cell by assuming that all base stations are transmitting with a preassigned power and feeds back the selected cell with its SINR.
– Each base station allocates its resources to the best user selected among those competing for that cell. When user allocation is decentralized, two points need to be remarked.

The first one is related to the update of $\boldsymbol{\mu}^i$. This can be performed at the base stations, if users are served by only one base station, or it can be performed by the users after that the information on the resource allocation is sent to them. The second one is related to the evaluation of the user rates. This can be done based on the SINR evaluated by the user, which does not take into account the powers actually allocated to interfering users, because they are not known. Therefore, the allocated rate is not the actual rate supported by the transmission leading to possible outages. This can be avoided only by evaluating the SINR by using the worst-case values of interfering power in the vector $\mathbf{V}_m$, which can be further constrained to be less than a maximum value $P_{max}$ on each resource unit.

To counteract the losses that opportunistic schemes present when the number of users is moderate or low[4], while still preserving a decentralized implementation, the "power planning" concept (as [21] for time-division multiple access systems) can be introduced to preassign suitable power values to vector $\mathbf{V}_m$ with the additional constraint $p_{m,q} \leq v_{m,q}$.

---

[4] When the number of users is not very large, sum rate is maximized by allocating a number of users on each subcarrier and slot usually smaller than $Q$.

## 5 Conclusions

Multi-cell scenarios are present in a very large number of standards and systems. The trends of technology and the increased requirements in bandwidth usage efficiency dictate the tight re-use of the frequency bands in neighboring cells. To do that, effective interference management schemes are required to regulate optimally the transmitted power in each subcarrier in all cells. In this context, power planning was presented as a suitable tool to extract effective network parameters and requirements. Additionally, layered and distributed dynamic resource allocation algorithms were introduced in those scenarios that have predefined rate requirements and power may be adjusted to guarantee the provided QoS. Finally, cross-laying for multi-cell user scheduling is focused by Lagrangian duality to solve the same problem and ensure the QoS constraints.